\documentclass[manuscript]{aastex}     

\slugcomment{Submitted to AJ}

\shorttitle{Dwarf Novae in LMC}
\shortauthors{Shara et al.}

\begin{document}


\title{Erupting Dwarf Novae in the Large Magellanic Cloud}


\author{Michael M. Shara\altaffilmark{1}}
\affil{American Museum of Natural History, 79th St. and Central Park West, New York, NY, 10024}

\email{mshara@amnh.org}

\author{Sasha Hinkley}
\affil{American Museum of Natural History, 79th St. and Central Park West, New York, NY, 10024}

\email{shinkley@amnh.org}

\and

\author{David R. Zurek\altaffilmark{1}}
\affil{American Museum of Natural History, 79th St. and Central Park West, New York, NY, 10024}

\email{dzurek@amnh.org}


\altaffiltext{1}{Guest observer at the Cerro Tololo InterAmerican 
Observatory, which is operated by the Association of Universities for 
Research in Astronomy under cooperative agreement with 
the National Science Foundation. 
}


\begin{abstract}

We report the first likely detections of erupting Dwarf Novae (DN) in an 
external galaxy: the Large Magellanic Cloud. Six candidates were 
isolated from approximately a million stars observed every second night 
over 11 nights with the CTIO 8K $\times$ 8K Mosaic2 CCD imager. Artificial 
dwarf nova and completeness tests suggest that we are seeing only the 
brightest of the LMC DN, probably SS Cygni-like CVs, but possibly SU UMa-type 
cataclysmics undergoing superoutbursts. We derive crude but useful limits on the 
LMC DN surface density, and on the number of DN in the LMC. Many thousands 
of cataclysmic variables in the Magellanic Clouds can be discovered and 
characterized with 8 meter class telescopes.

\end{abstract}




\section{Introduction}

The study of cataclysmic variables (CVs) in our Galaxy is plagued by the 
same problem that afflicts so many other areas of astrophysics: 
uncertainty in the distances to, and hence luminosities of the objects 
being studied. A second and no less severe problem is the space density 
and spatial distribution uncertainties caused by our parochial view of 
the Milky Way. While we can detect erupting Galactic classical novae 
several kpc from the Sun, most other CVs are located closer than about 
500 pc. The reason is simple: novae often achieve 
$10^5\,$ $L_{\odot}$, rivaling the most luminous objects in our Galaxy 
for weeks at a time, while dwarf novae (DN), nova-like variables and 
their magnetic cousins rarely exceed 10 $L_{\odot}$. Astronomers have 
succeeded in cataloguing barely a thousand Galactic CVs in over a 
century of searching. Only for a handful are ironclad distances 
published from Hubble Space Telescope parallaxes \citep{har99,mca99, mca01}. Expansion 
parallaxes for classical novae contribute another 20 or so reasonably secure distances 
and luminosities (see \citet{war95} for a summary).

It would clearly be of enormous benefit to CV science if hundreds or 
thousands of cataclysmic variables, all at the same distance, could be located. 
Accurate luminosity functions, and bias-free period distributions and 
eruption frequencies would become available to confront 
inadequately-constrained theoretical models. New sub-classes of CVs 
might be discovered, and systematic variations in outburst properties 
and binary orbital distributions would likely be uncovered.

The situation is improving for CVs in a few special places: the crowded 
cores of globular clusters. Recent detections of dozens of CV candidates 
in 47 Tuc \citep{gri01,kni02}, allow, for the first time, the 
comparative study of many cataclysmics, {\it all at the same distance}. 
Unfortunately, HST and Chandra are essential for followup studies, and 
these telescopes are two of the rarest commodities in astrophysics. It 
may be decades before the newly discovered CVs are fully characterized. 
Worse, many of these CVs probably formed by tidal captures and/or 
evolved violently under the influences of passing stars. Only one 
globular DN is easily resolved and studied from the ground: V101 in M5 
\citep{mar81,sha90}. Its rather long (5.79 hour) period 
\citep{nei02} warns us that it may be anomalous. CVs in the cores of 
globular clusters have much to teach us, but they're likely to be very 
different from field CVs.

The outlook is potentially more promising in the Magellanic Clouds.
As in the case of globular clusters, all CVs discovered in the Clouds 
are at nearly the same distance, so direct luminosity comparisons are 
meaningful. Spatial densities of field LMC and SMC stars are roughly 
comparable to those in the neighborhood of the Sun, rather than those 
typical of globular cluster cores, so tidal encounters almost never 
happen. CVs in the Clouds must form and evolve via ordinary binary 
evolution, just as in the field in our own Galaxy. It is certainly true 
that the metallicities of the red dwarf companions to the white dwarfs 
in Magellanic CVs will usually be lower than those in the Galaxy. Lower 
metallicity is a much less drastic effect \citep{ste97} than dynamical 
interactions \citep{hur02} in globular clusters, so direct comparisons 
between Galactic and Magellanic CVs should be extremely fruitful. 
Finally, the Clouds must be home to $10^6\,$ or more CVs (see below), an 
enormous sample likely to contain every subtype of CV we know, and 
perhaps some that we don't yet recognize.

Before carrying out Galaxy-LMC population comparisons, we must, of 
course find CVs in the Clouds. Since 1897 about 35 erupting classical 
novae have been spotted in the LMC and SMC. Accurate (1-2 arcsecond) 
positions for most of these, and recovery of some in deep $U$, $B$ and $V$ 
images has recently occurred \citep{sha04}. The current sample of 
quiescent Cloud novae is important to follow-up, but will grow in number 
only very slowly. This is because all classical novae must recur 
\citep{for78} with inter-eruption times of at least $10^4\,$ yr. Thus 
the many thousands of classical novae that exist in the Clouds will only 
slowly reveal themselves, via eruptions, over many millennia.

As dwarf novae are a substantial sub-population of CVs which reveal 
themselves through outbursts every few weeks to months, {\it almost all} 
DN in any given field should be identifiable, at least in principle, in 
a deep survey of order 6-12 months in length. Luminous erupting DN 
should achieve $U\sim 22.0$, $V\sim 22.5$ near maximum (see section 6) and thus 
be detectable except in the most crowded LMC fields.

Surveying the entire LMC and SMC often enough to find nearly all 
erupting DN is a daunting observational program that can and will 
eventually be undertaken. In this paper we report a much more modest but 
realistic feasibility study to demonstrate, for the first time, the 
existence of erupting DN in the LMC. We also derive a crude estimate of 
the total number of luminous DN in the LMC.

In section 2 we present the observational strategy and database, while 
the photometry calibrations are described in section 3. Completeness 
tests are detailed in section 4, and the six erupting DN candidates are 
shown in section 5. Simulations of erupting DN in the LMC are compared 
to these candidates in section 6. We discuss whether the candidates 
could be other kinds of variables, and the implications for CV numbers 
and space densities if the candidates are true DN in section 7. We 
briefly summarize our results in section 8.

\section{Observations}

To maximize our chances of detecting erupting LMC DN we requested the 
longest dark run with the largest telescope (the 4-m)and widest field 
imager (Mosaic2 8K $\times$ 8K) available in 1999 at CTIO. We were awarded 
the dark nights of 1999 December 2, 4, 6, 8, 10 and 12. All nights were 
clear, with seeing mostly in the range $1.0 - 1.8$ arcsec. Because DN are 
usually remarkably blue ($U - V \sim -0.7$ is a good rule-of-thumb; though 
much redder values do occur \citep{bai80}) we opted to image our chosen 
LMC field in $U$ and $V$ filters. Three to eight images were obtained each 
night in each filter, with total exposure times (typically) of 2.5 hours 
in $U$ and 1 hour in $V$. A detailed log is given in Table 1. The plate 
scale was 0.270 arcsec/pixel, and our total field of view was 1362 
arcmin$^2$ = 0.38 deg$^2$.

The LMC field we chose is centered at RA (J2000) 05:33:36.7
and DEC (J2000) -70:33:44. The field of view was chosen because five erupting 
classical novae (the novae of 1948, 1970a, 1970b, 1981 and 1988a) have been observed 
in the area covered by our CCD array. . . 
more than any other location in or near the LMC. We were thus able to monitor 
these five old novae ``for free'' while searching for erupting dwarf novae. The results of the 
old nova monitoring will be reported elsewhere \citep{sha04}.
An overall view of the location of this field within the LMC is shown in Figure 1, 
with a magnified view of the eight CCD fields superposed. The field is at the 
southeastern end of the LMC bar, and crowding is significant. The large surface 
density of stars hampered efforts to detect faint DN, but was critical to maximize the 
chances of detecting at least a few bright erupting DN.

The individual Mosaic2 $2048 \times 4096$ images were combined using the ``Montage2'' 
routine contained within the stand-alone DAOphot package. On a given night, 
individual frames were taken with a total vertical
dither of $\sim 10$ pixels. These frames were run through DAOphot's
matching program ``DAOmaster'' to derive the subpixel frame-to-frame
shifts. These shifts were passed to Montage2, which produced a sky
subtracted median image of the individual frames.  

\section{Photometry Calibration}

The calibrations were derived from a set of standard magnitudes taken from
the central region of M67 as given in \citet{mon93}. To ensure that the central 
region of M67 was observed on all eight chips, the cluster was observed over 
two nights on the ``right hand'' side of the array and then later on the ``left 
hand'' side. However, the ``left hand'' side was observed on the fifth night, 
and was therefore unusable due to poor seeing. Nevertheless, the zero-points
between chips were consistent to within .05 magnitudes, and the zero-points
derived for the ``right hand'' side were used for the entire data set.

Due to the variable seeing, there also exist very subtle differences in
the photometry on a night-to-night basis. Thus, in addition to the global
zero-point calibration uncertainty described above, there also exist night-to-night zero-point corrections on the order of .05-.1 mag. These corrections were
found by first searching for the least variable (most constant) stars
that returned valid photometry for all of the six nights. This was done in
an iterative manner to locate stars with variability at least three times less 
than the average of the ensemble. Once this goal was achieved, a night-to-night 
average was calculated, and this was subtracted from the photometry of the 
returned candidates. None of these rather small uncertainties has any effect 
on the candidate detections and conclusions of this paper.

\section{Completeness tests}

Accurate photometry and astrometry of any source is limited not only
by the source's brightness, but also by the degree of crowding in the
field. Simulations were performed to determine the efficiency of recovery of a 
collection of synthetic stars within a representative sub-region of the dataset 
as a function of brightness. These simulations use the actual point spread 
function for a given night and filter, yielding a measure of the effective 
plate limit and the overall detection sensitivity of these observations and
data reductions.

First, the same small sub-region of the entire field was chosen in both the
$U$ and $V$ bands and from each of the six nights. The stars in this 500
$\times$ 500 pixel sub-image showed crowding typical of the rest of the
dataset ($\sim$ 4000 stars in the sub-image). Moreover, the star brightness 
range in the overall dataset was well matched by the range in this sub-image.

Next, in order to generate a set of synthetic stars, the ``addstar''
routine contained within DAOphot was used to place 100 copies of
the point-spread function randomly across an image and at a set
brightness. This was repeated 10 times, for a total of 1000 synthetic stars at 
a given brightness, night, and filter. Then ``allstar'' returned photometry on
all the stars from these 10 runs. The fraction of the input synthetic stars that
were recovered gives the level of completeness for that given
brightness. A synthetic star was considered ``recovered''
if it was measured within $\pm 0.5$ magnitudes of its input value. The entire
procedure was then repeated with a different brightness for
the input synthetic stars. A total magnitude range of $\sim18.5 - 24.0$
for both the $U$ and $V$ filters was covered in increments of 0.25
magnitudes each. The results are shown in Figures 2a and 2b. 

Each of the completeness curves reflect well the seeing of each night. For 
example, the completeness for the third night (best seeing) is 70\% at $U=22$, 
and 40\% at $V=23$, typical of an LMC DN near eruption maximum. On the fifth 
night (the worst seeing night) the completeness at these same magnitudes is
barely 2-3\%. Fortunately every night's imagery except that of night 5 was deep enough 
and complete enough for us to expect to see at least $\sim 10-20\%$ 
of all erupting DN at or near peak brightness. 

\section{Candidate Dwarf Novae}

For a field with such a high stellar density ($\sim1000$
stars/arcmin$^2$), the construction of a good point spread funtion (PSF)
is hampered most by neighboring stars crowding the PSF stars. Much effort was 
expended to automate the search for variables entirely, comparing lists of PSF
photometry stars on successive nights. Night-to-night seeing changes produced 
many candidates in the automated search that were rejected upon visual inspection.
Rejection of a candidate invariably occurred because the ``candidate'' revealed itself 
to be an artifact created by overlapping PSFs of several stars, changing with the seeing
from night to night.

The dwarf-nova candidates were ultimately found by blinking rapidly through the 
best four of the six nightly frames to look for any subtle changes in the 
appearance of the stellar field. Dwarf novae should rise from invisibility to 
easily detectable in the 48 hours between observations. To look for changes on 
a fairly detailed level, each chip was visually divided into 32 sub-regions, 
and blinked rapidly. Brightening from previously empty regions of the sky is 
easily seen. A star that is ramping up or down in brightness by $0.5-1.0$ mags 
can also be easily detected since the transition between the first and last 
night is very prominent. Artificial dwarf novae placed in the frames were 
visually detected with efficiencies similar to the completeness curves of Figure 2.  

To be recognized as a candidate, a variable had to be visible in both the $U$ and 
$V$ medianed frames of at least one of the six nights, and to have varied by
over one magnitude in both filters between any two nights. This selection method 
initially revealed 14 dwarf nova candidates from the entire dataset. A second 
criterion was then applied to weed out short-period, large amplitude variables: 
no candidate could vary significantly (more than about 1.0 magnitudes) from 
frame to frame during any one night. It is certainly true that the light output 
of many CVs ``flickers'' on timescales of seconds to hours, and that some CVs 
undergo short, deep eclipses. We nevertheless adopted this conservative approach 
to eliminate non-CV, short-period eclipsing binaries. Inspection of each night's 
individual images confirmed that 7 candidates were likely eclipsing binaries, 
and one candidate was so crowded that we cannot say for sure that it is a real variable.
The eclipsing variables may be useful for LMC distance determinations or other projects, 
and so we supply their finder charts, and coordinates in Figure 8 and Table 3, respectively. 

We are thus left with six good candidate erupting DN. We define a good candidate 
as one that is seen in both $U$ and $V$ on at least one night; on
all images in each filter of that night; and whose variability characteristics 
are consistent with those of at least one well-studied Galactic dwarf nova (see 
next section). The nightly images of the six DN candidates are shown in Figure 
3, and their photometry is presented in Figure 4. 
The poisson error is the dominant photometric error in these plots, although readnoise 
and sky noise also have been incorporated into the error bars shown.  
We defer a discussion of these candidates and their implications until after the next 
section, where we simulate expected DN images and light curves.

\section{Simulations of Known Dwarf Novae}

To complement the completeness tests described in section 4, simulations of 
eruptions of several well known and characterized Galactic dwarf novae, 
artificially placed in the actual LMC data, were needed. Eruptions of SS Cyg, 
U Gem, SS Aur, AR And, and EM Cyg were all simulated in a relatively uncrowded
but otherwise representative patch of the LMC field (An uncrowded field was chosen to 
highlight the photometric appearances of LMC DN, unhindered by the crowding 
or incompleteness addressed in Figure 2). The images from these simulations are shown in Figure 5.

The three dwarf novae SS Cyg, U Gem and SS Aur have accurately determined
parallaxes \citep{har99} and, using their well tabulated apparent magnitudes 
\citep{war95}, the absolute magnitudes are easily calculated. The absolute 
magnitudes for AR And and EM Cygni were obtained from \citet{war95}. These five 
objects were chosen to cover the range of absolute magnitudes seen for erupting 
and quiescent DN. Our simulated dwarf novae all reflect the rise, plateau, and 
decline times from \citet{szk84}, start at quiescence on the first night of 
observations and attain maximum brightness on the second observing night (48 
hours later), have all been placed at the distance of the LMC and, with one 
exception noted below, use mean $(U - V)$ colors from \citet{bru94}. We have 
assumed that the LMC DN are dimmed by $A_V$ = 0.3 and $A_U$ = 0.5, in accord 
with \citet{cle03}. PSF photometry was performed on these frames; the resulting 
light curves and retrieved photometry are shown in Figure 6. 

An additional simulated SS Cygni eruption sequence (labeled ``SS Cyg sequence 2'' 
in Figure 6) incorporated more detailed information about this object's color 
evolution with time. During its eruption, SS Cygni becomes distinctly more red, 
reaching $(U - V) \sim 0.2$ before attaining maximum brightness \citet{bai80}. This second 
SS Cyg sequence utilizes these colors, and has been shifted by one day in phase. 
That is, the first simulated observation of this second sequence catches the dwarf nova 
on the rise. (``SS Cygni sequence 1'' shows the artificial DN in quiescence on the 
first night). This second SS Cyg sequence was included because it mimics rather well the overall 
observed light curves of DN candidates SHZ1, SHZ4 and SHZ6, though the latter 
two objects are still redder than the simulated DN. We regard this difference 
as minor because reddening does vary strongly, and on small spatial scales, 
across the LMC \citep{goc02}. 

It is clear from Figure 5 that we are unlikely to have been able to detect erupting DN like U Gem, SS Aur, 
AR And or EM Cyg--all are too intrinsically faint to be detected, in our crowded fields. If our candidates are 
really DN they are much more likely to be SS Cygni, or possibly SU UMa-type CVs in superoutburst. 

\section{Interpretation}

\subsection{Have We Found Erupting LMC Dwarf Novae?}

Comparison of the six candidates shown in Figure 3 with the
LMC-DN simulation images shown in Figure 5 (particularly ``SS Cyg 
Sequence 2'' and ``SU UMa superoutburst'')
demonstrates that the brightness and variability behaviors of our 
candidates are broadly consistent with those expected of luminous erupting dwarf 
novae in the LMC. This is further supported by comparison of the 
measured and simulated light curves of Figures 4 and 6, respectively. 
However, it is by no means certain that all or even some of the six 
variables shown in Figure 3 are really LMC erupting dwarf novae.

What are the other possible variables that might mimic LMC DN behavior? 
Amongst these are: chance superpositions of background supernovae
or classical novae, gamma ray bursts (GRB), microlensing events, Milky 
Way variables along the line of sight to the LMC, and non-CV LMC variables. 
The microlensing hypothesis can immediately be discarded because of 
the non-grey light curve behaviors, non-symmetric shapes of the light 
curves and (in most cases) overly long time at maximum light.

GRB afterglows typically achieve R or I $\sim 20$ one day after 
outburst. While the peak brightness of these variables is in accord with 
the GRB hypothesis,
we can eliminate this possibility because none of the candidates 
declines quickly enough to be a GRB. Large area, multi-epoch surveys for 
faint variables (e.g. \citet{haw87}) show that the surface density of 
our variables is far too high for them to be field RR Lyraes, classical 
novae or supernovae.

The brightness and moderately blue colors of the candidates rule out other
types of Galactic variables. RR Lyraes in the Galaxy or LMC would be 
considerably brighter, and flare stars don't match the observed nightly 
brightness profiles and/or blue colors. 

A final possibility to consider is that some of our candidates might be 
Galactic or LMC eclipsing binaries. Galactic binaries would have to be rather
low luminosity systems (say $M \sim 10 - 15$) and thus much redder than 
observed to appear at $m \sim 20 - 22$. Eclipsing LMC systems would range in 
luminosity from $M \sim 0$ (SHZ6) to $M \sim 2.5$ (SHZ3 and SHZ5), corresponding 
to A-type main sequence stars. The amplitudes of light variation in SHZ1, SHZ3, SHZ4, SHZ5 and SHZ6
are greater than 0.7 magnitude and thus preclude equal mass (and brightness) binaries, but a hot 
pre-white dwarf + a cooler companion star could mimic the observed light curves for SHZ1, SHZ5 and SHZ6. 


As noted earlier, we see no sign of significant variability for any of 
the six candidates on the individual frames taken during the two to five 
hours of observations during each of the six nights of the run. The 
candidates' light curves support their tentative identifications as DN, though some may turn out 
to be other kinds of variables. 

To be absolutely certain of these objects' identities will require 
challenging follow-up observations. Four possible confirmation 
techniques are the following:

1) Spectra near quiescence (to demonstrate the presence of Balmer 
emission lines) would be definitive proof, but the DN are then expected 
to be near 25th magnitude...beyond the likely capability of even an 8 
meter telescope in such crowded fields.

2) Imagery every night or two for several months with a 4 meter class 
telescope should reveal repeated eruptions separated by weeks to months. 
Such a program would also distinguish erupting CVs from LMC eclipsing binaries, 
where the emergence of a hot, blue star from eclipse---seen only at one epoch---can be 
confused with a genuine DN eruption. This can be done, but will be very 
demanding of large telescope time.

3) Ultraviolet imagery of the fields of the six candidate dwarf novae 
with the Hubble Space Telescope (HST) might reveal UV-bright objects, as 
the spectral energy distributions of most CVs rise sharply into the UV. 

4) Several hours of time-resolved UV or optical photometry, again with 
HST, might reveal the flickering and/or orbital modulation 
characteristic of CVs.

\subsection{Expected Dwarf Nova Populations in the LMC}

Space density estimates of CVs near the Sun suggest 
$10^{-5}\,$stars$\,{\rm pc}^{-3}$ \citep{pat98}. Space density estimates 
of all types of stars in the solar neighborhood yield 
$10^{-1}\,$stars$\,{\rm pc}^{-3}$, suggesting that about 1 CV exists in 
the Galaxy - and probably in the LMC - for every $10^4\,$ stars. The LMC 
displays a V-band luminosity of -18.1, corresponding to a mass of 
$10^{10}\,$ $M_{\odot}$ \citep{cox99} and a population of a few times 
$10^{10}\,$ stars.

If the Galaxy and LMC manufacture CVs with similar efficiencies and 
rates then we estimate a total LMC CV population today of a few times 
$10^6\,$ objects.

About half of known CVs are DN \citep{dow01} which suggests a total LMC
DN population of order $10^6\,$. These DN are spread across the 10 
$\times$ 10 degree surface of the LMC. Erupting classical novae are 
observed to be distributed quite uniformly across the face of the LMC 
\citep{van88}, consistent with them belonging to an old population. A 
surface density of 10,000 DN per square degree across the LMC is thus 
expected.

\subsection{Detections versus Expected Detections}

The size of our field of view was 1362 arcmin$^2\,$ = 0.38 deg$^2\,$, 
which should include 0.38 $\times$ 10,000 = 3,800 DN. The average time 
between eruptions for 21 well studied Galactic DN is 29 days 
\citep{szk84}. The length of our observing run (11 nights) suggests that 
any DN erupting on the nine nights between nights 2 and 10, inclusive, 
could have been detected (if it became sufficiently luminous). Hence 
(9/29) $\times$ 3800 $\sim$ 1200 DN eruptions should have occurred in our field of view 
that were (at least in principle) detectable.

The 4 meter telescope seeing varied significantly over the six nights on 
which we observed, as clearly seen in the completeness curves of Figure 
2. Only on the second, third and sixth nights were conditions good 
enough to allow straightforward detection (with 20\% to 40\% 
completeness due to crowding) of DN reaching $U\sim 22-23$. Assuming 
(conservatively) that we missed half of all DN because only 3 of 6 
observing nights were ``good'', and that we missed 80\% of all
eruptions even on the three ``good'' nights because of crowding, we might 
still have expected to have detected $0.2 \times 0.5 \times 1200 = 120$ 
erupting DN.

The apparent brightness of a dwarf nova depends both on the underlying binary system
inclination and orbital period. Nearly face-on disks and longer orbital periods (which 
have larger and brighter disks) will dominate a magnitude limited sample \citep{pac80, war86}. 
High inclination systems will be up to 2 magnitudes fainter than those seen nearly face-on, 
and will therefore be lost in a sample that is detecting only the brightest tip of 
the distribution. 

Longer period systems will have larger and brighter disks, and thus might be expected to 
dominate our sample. However, the number of short-period DN ($P < 2$ hr ) is twice the number of
long-period systems ($P > 3$ hr), and it is these short-period 
systems that undergo SU UMa-like superoutbursts. SU UMa superoutbursts typically last $\sim 3$ 
weeks, and system brightnesses might match those observed for our six candidates. This will be 
clearer when parallaxes are available for at least a few SU UMa CVs. 

The fact that we detected at most six DN when (an admittedly simple) 
model predicts 120 suggests that many erupting systems were too faint to be 
detected. This is in accord with the model light curves of Figure 6, and 
reinforced by the simulated images of Figure 5 showing that DN like U 
Gem, SS Aur and EM Cyg would all have been missed in our survey, and even (relatively 
luminous) AR And would have been possible but challenging to detect. What fraction of 
DN are at least as luminous as SS Cyg or SU Uma at maximum? Unfortunately, 
the answer is unknown, because the distances to all but four Galactic DN are 
too uncertain to yield a believable luminosity distribution. (We note that the 
canonical literature distance to the prototypical DN, SS Cyg, was in error by 
a factor of two until HST parallaxes became available in 2000). If our six 
candidates are all bona fide erupting DN in the LMC (of either SS Cygni or SU UMa type) 
this suggests that 95\% of all DN are less luminous than these prototypical dwarf novae. 
This percentage rises if fewer than six of our candidates are really DN.

\section{Conclusions}

We have carried out an 11 day-long survey for erupting dwarf novae in 
the Large Magellanic Cloud, during which we imaged 0.38 deg$^2\,$ in $U$ 
and $V$ filters on alternate nights. Artificial DN and completeness tests 
confirm that we imaged faint enough (to 23rd mag on the better nights) 
to detect SS Cygni-like eruptions. Six candidates were isolated from 
approximately a million stars observed with the CTIO 8K $\times$ 8K 
Mosaic2 CCD imager. If these are true LMC DN, then they are amongst the most luminous of 
outbursting dwarf novae in the Clouds; SS Cygni-like outbursts or possibly SU UMa-like 
superoutbursts. We estimate the total LMC DN population as $10^6$, 
and the LMC surface density as $10^4$ DN/deg$^2$. This suggests that at least 95\% of erupting 
LMC DN are fainter than the six candidates we present, and were thus missed. 
An extra magnitude in sensitivity and 
improvements in spatial resolution will likely lead to the discovery of 
thousands of DN in the Magellanic Clouds.


\acknowledgments

We are indebted to the Cerro Tololo Observatory Directors Robert Williams and 
Malcolm Smith for the generous allocation of observing time for this project. We also 
thank Tony Moffat, Darragh O' Donohue, Brian Warner and an anonymous referee  
for valuable comments.

\appendix 

While the time coverage we have of the seven variables noted in Section 5 is far too sparse 
for any attempt at period determination, we include their coordinates in Table 3 and finder
charts (in $U$ and $V$, minimum and maximum) in Figures 7 and 8. 


\clearpage



\begin{figure}
\figurenum{1}
\caption{Field of the LMC imaged in our search for erupting dwarf novae, 
shown as an insert over the entire LMC. The $8^\circ \times 8^\circ$ image is 
an $R$-band MACHO image, while the insert is our $0.61^\circ \times 0.61^\circ$
CCD FOV imaged through a $V$-band filter. The scale bar on the bottom left 
is $1^\circ$ long. N is up, E is left in the image. The locations of the six erupting 
DN candidates are shown as numbered circles in the insert}
\end{figure}

\clearpage

\begin{figure}
\figurenum{2a}
\plotone{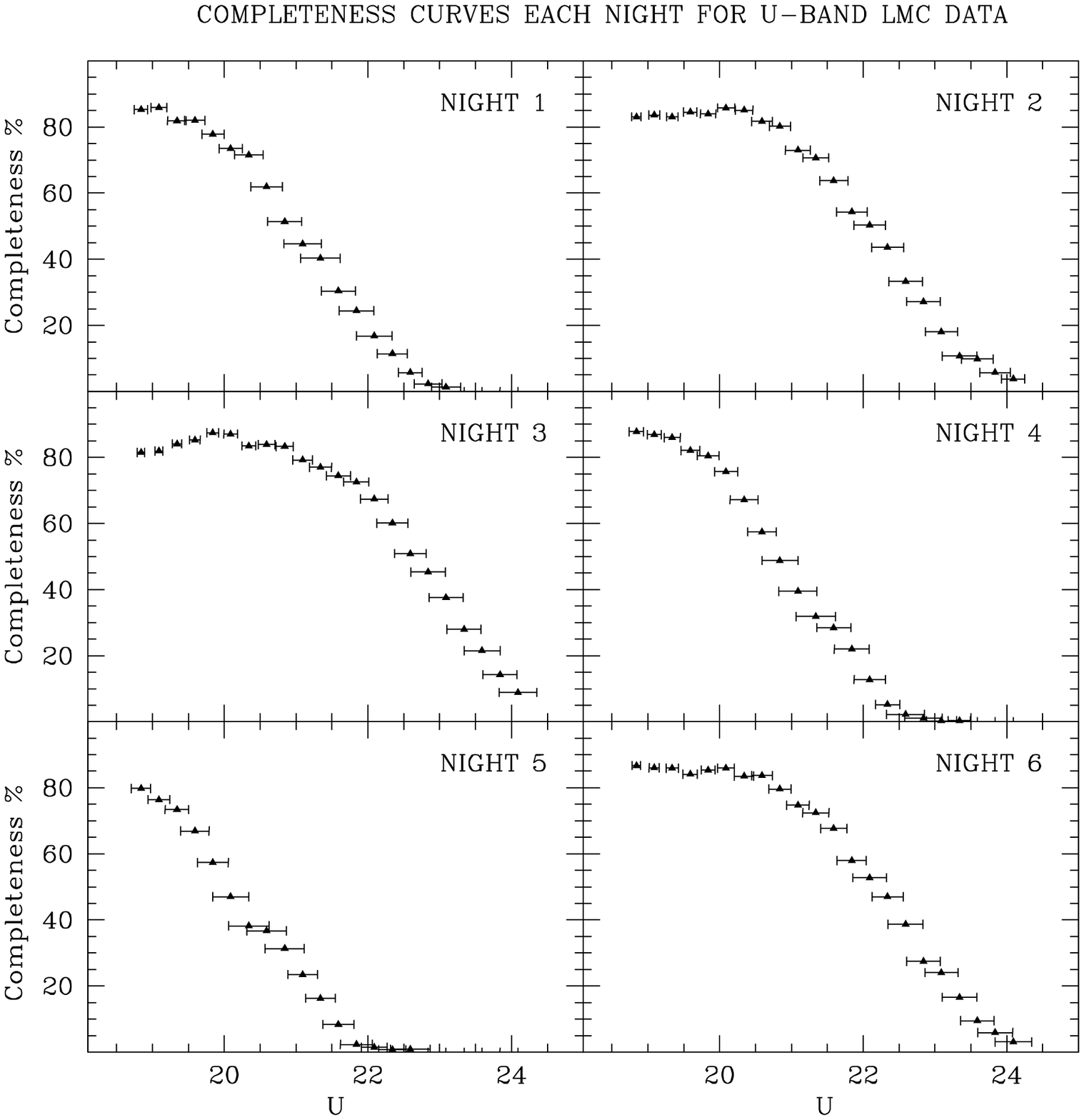}
\caption{Completeness curves for each of six nights of data in the $U$ 
band images. See text for details.}
\end{figure}

\clearpage

\begin{figure}
\figurenum{2b}
\plotone{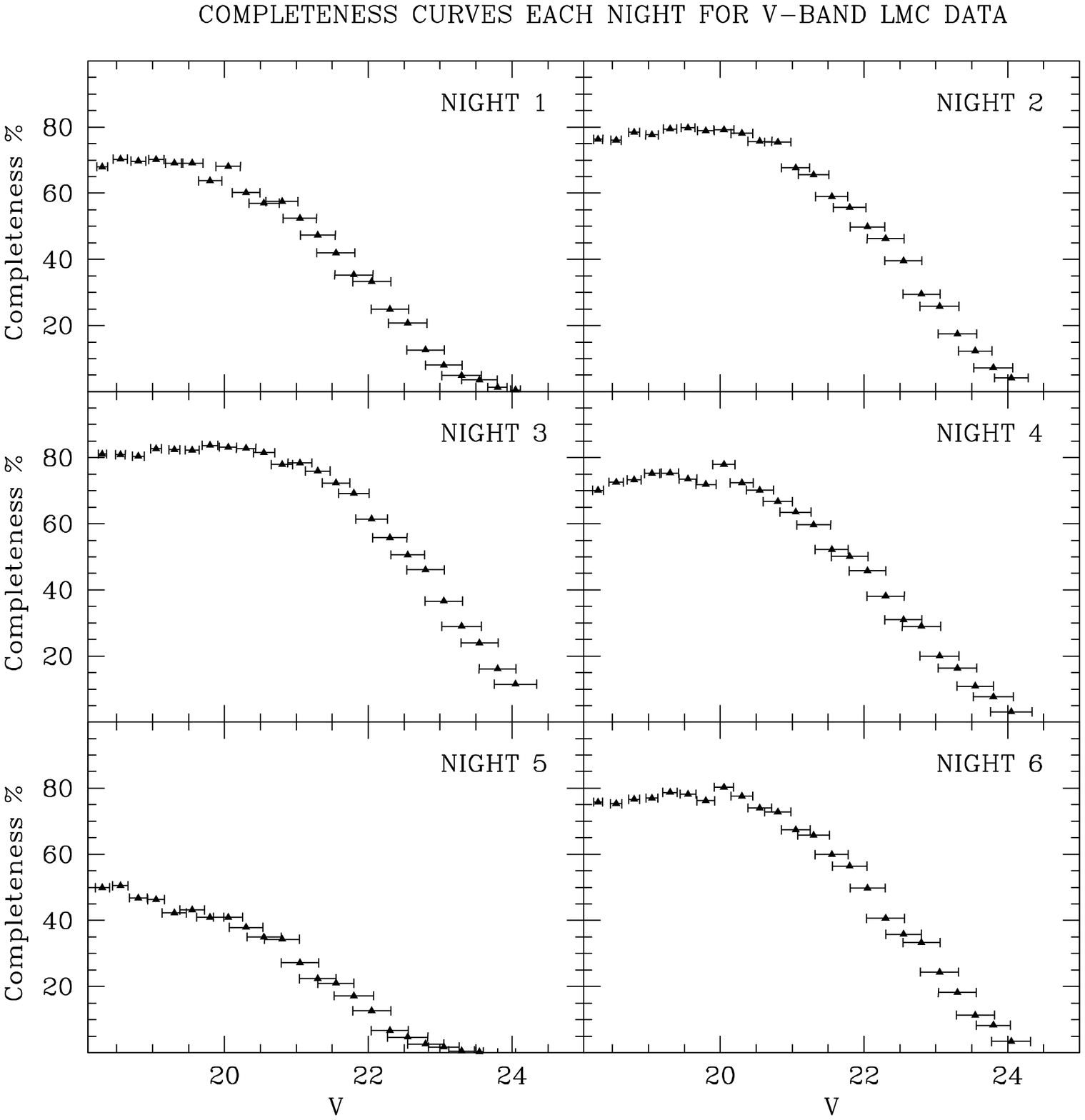}
\caption{ Completeness curves for each of six nights of data in the $V$ 
band images. See text for details.}
\end{figure}

\clearpage

\begin{figure}
\figurenum{3}
\plotone{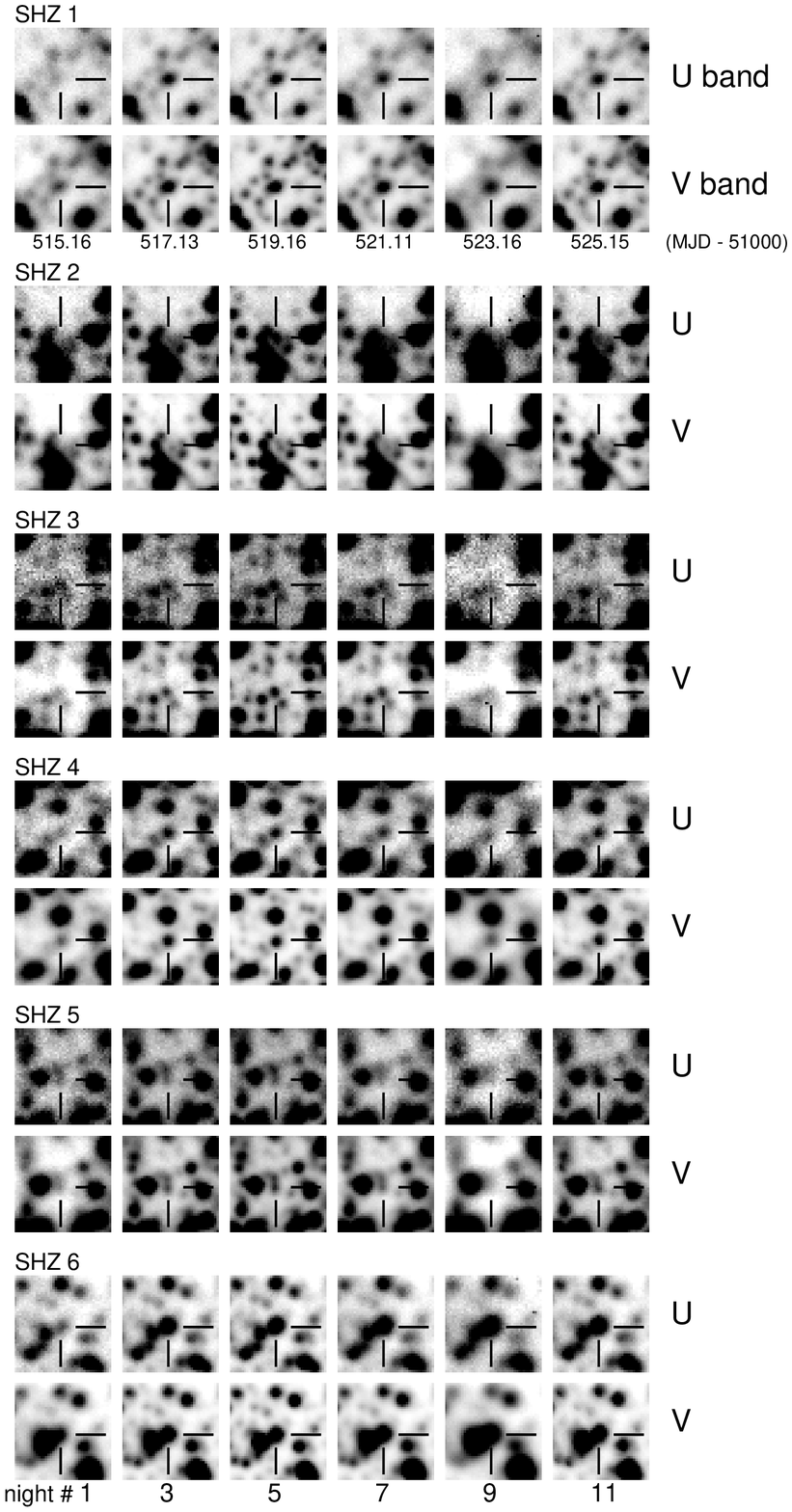}
\caption{Nightly medians of images in $U$ and $V$ bands of the six candidate 
erupting dwarf novae in the LMC.}
\end{figure}

\clearpage

\begin{figure}
\figurenum{4}
\plotone{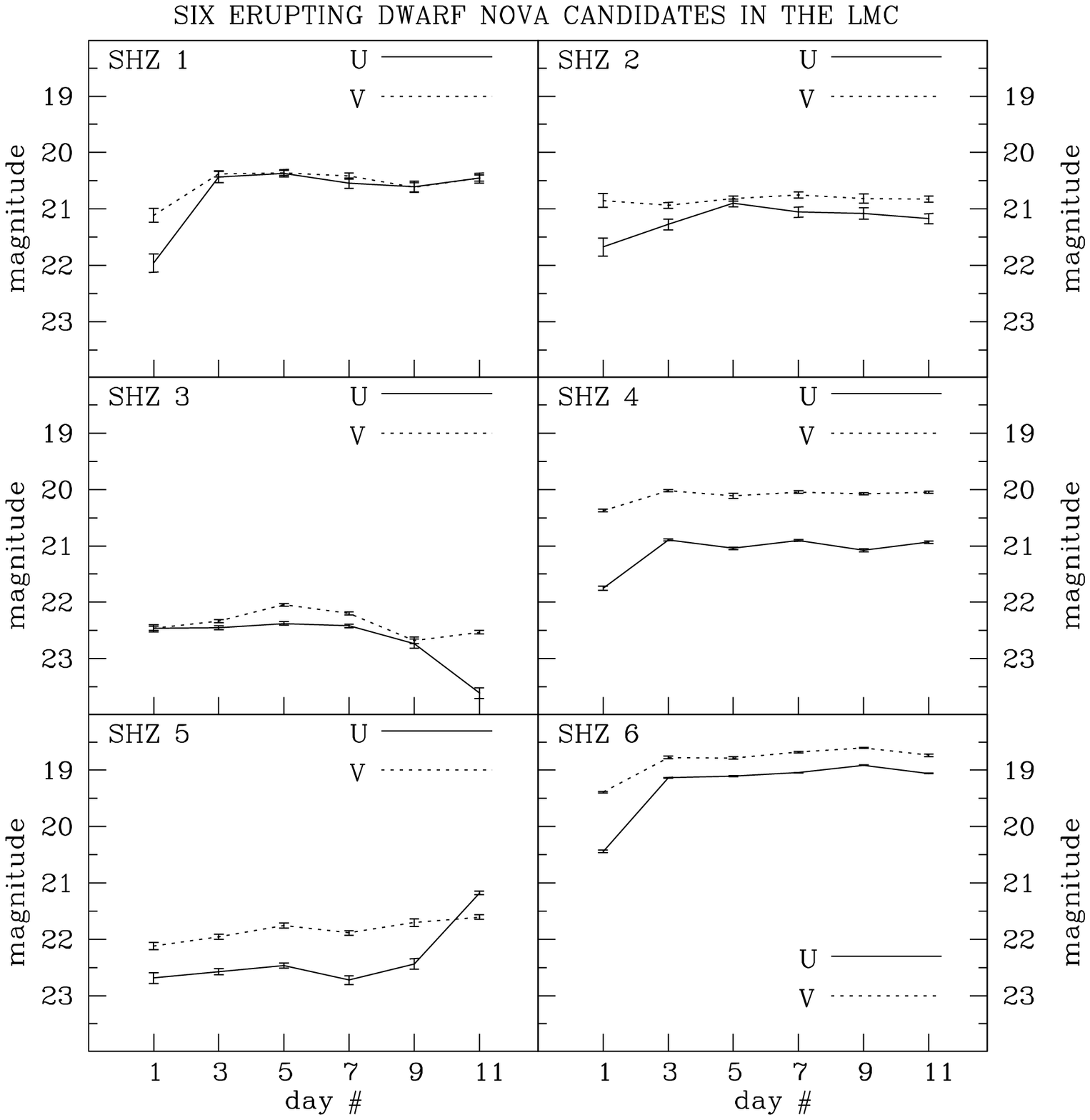}
\caption{$U$ and $V$ light curves of the six dwarf nova candidates shown in 
Figure 3.}
\label{sphot_fig}
\end{figure}

\clearpage

\begin{figure}
\figurenum{5}
\plotone{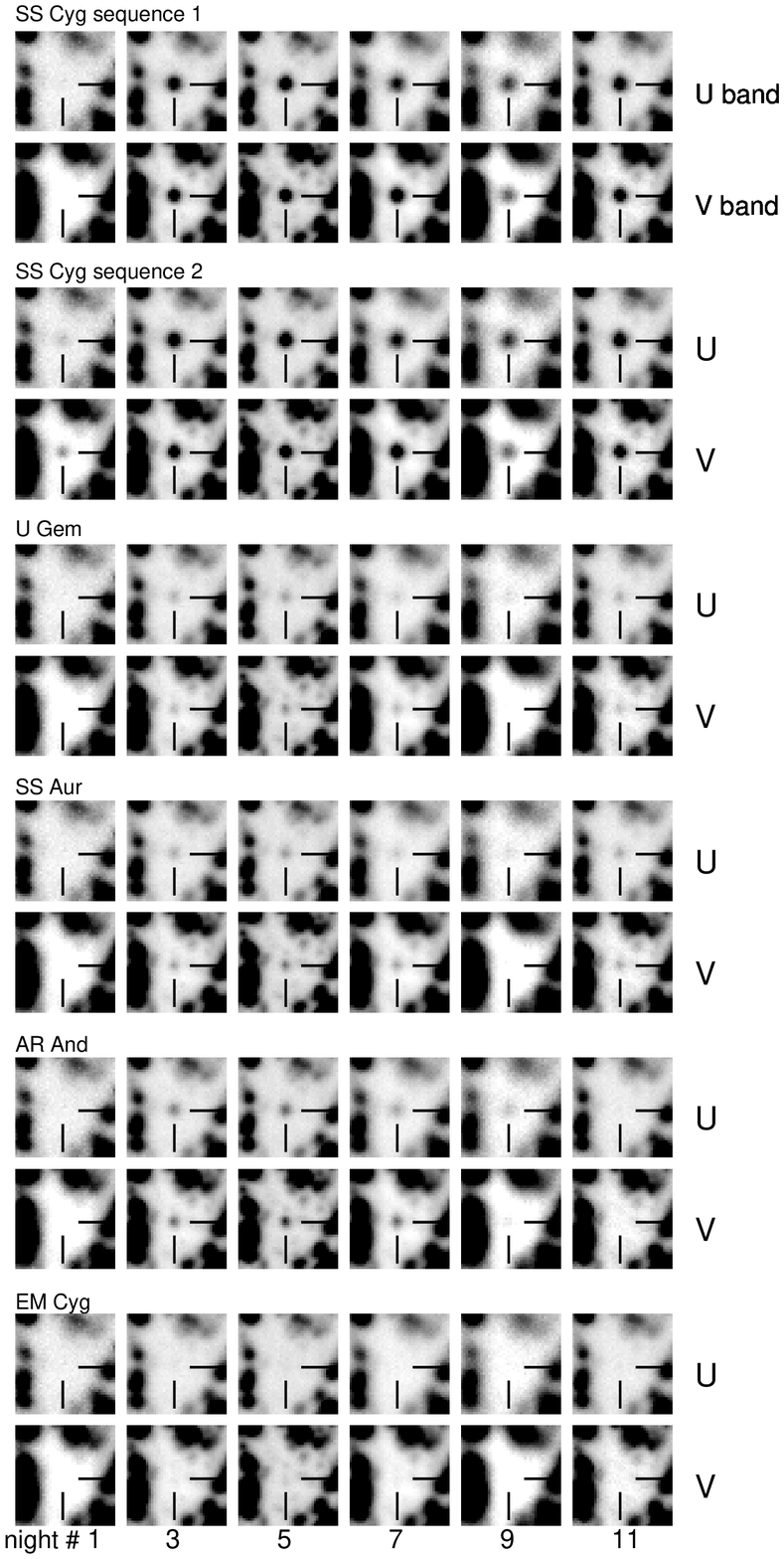}
\caption{Simulated images of erupting Galactic dwarf novae placed in the 
LMC. See text for details.}
\end{figure}

\clearpage

\begin{figure}
\figurenum{6}
\plotone{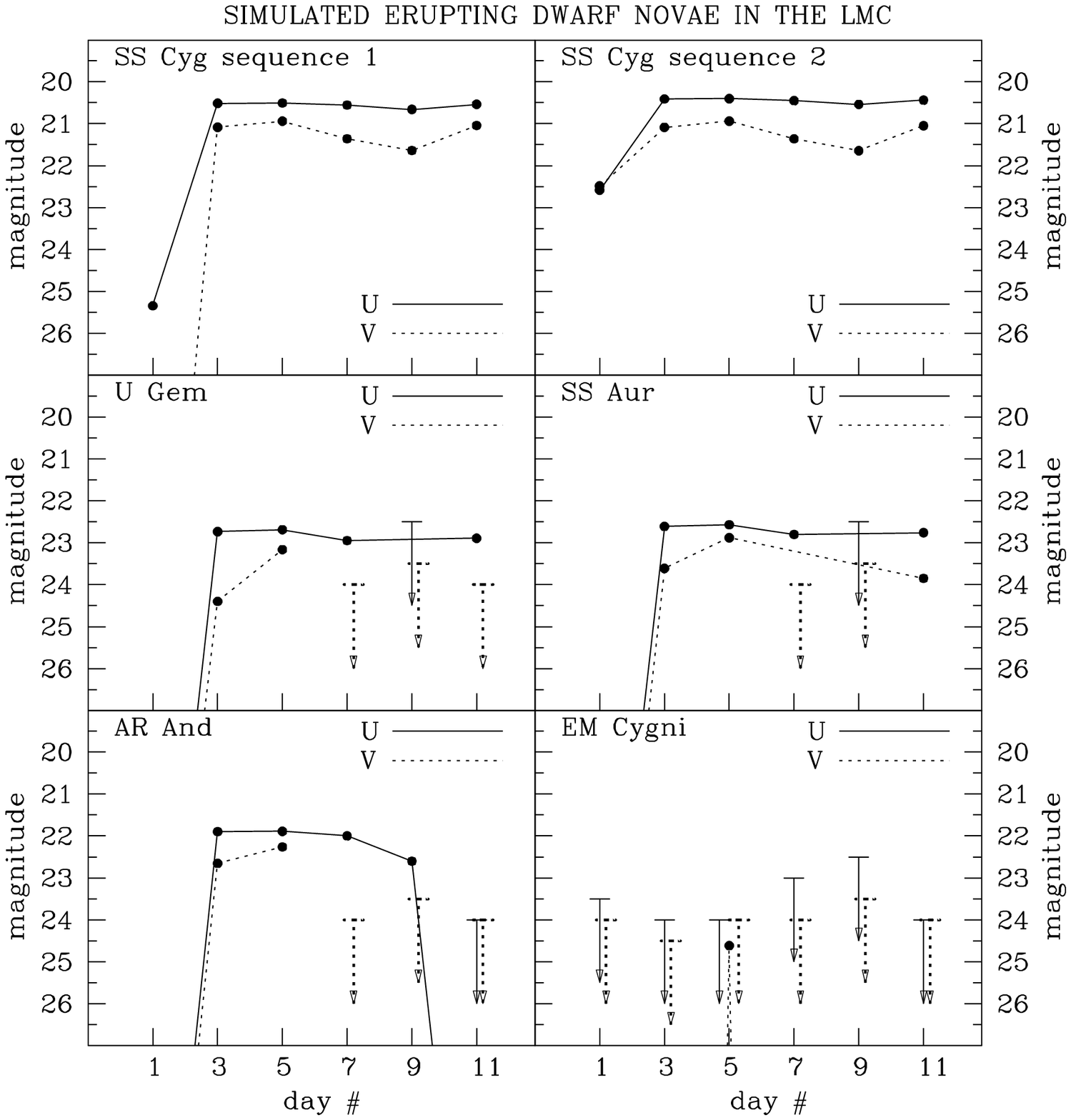}
\caption{Simulated light curves of the dwarf novae of Figure 5. The ``dips'' and magnitude
limits on night 9 are due to the particularly poor seeing of that epoch.}
\end{figure}

\clearpage

\begin{figure}
\figurenum{7}
\plotone{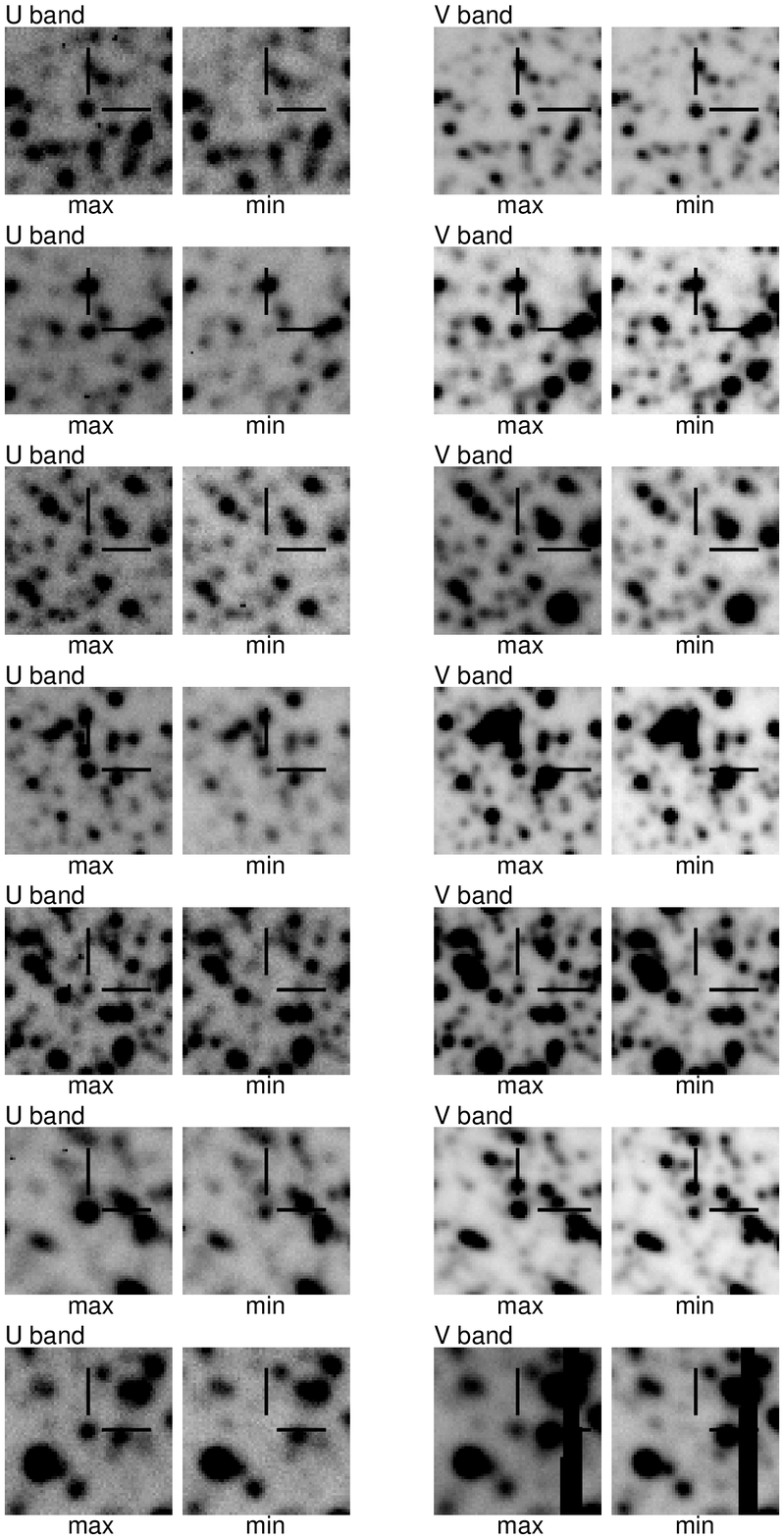}
\caption{Individual exposures in $U$ and $V$ bands of the seven 
eclipsing binaries/variables at their maximum and minimum brightnesses.}
\end{figure}

\clearpage

\begin{figure}
\figurenum{8}
\caption{Same as Figure 1, except the seven variables (probably eclipsing variables)
discussed in the text are shown as numbered circles, with prefix ``V'' in the insert}
\end{figure}

\clearpage

\clearpage
\begin{deluxetable}{lllllcc}
\tabletypesize{\scriptsize}
\tablecaption{CTIO 8K x 8K CCD Observations}
\tablewidth{0pt}
\tablehead{\colhead{UT Day} & \colhead{$t_{exp}(U)$ (min)} & \colhead{$t_{exp}(V)$ (min)} & \colhead{$U$ seeing} & \colhead{$V$ seeing } & \colhead{Completeness ($U=22$)} & \colhead{Completness ($V=23$)}}
\startdata
12.03.99 & $3\times30\rightarrow 90$   & $3\times10\rightarrow30$  &  1.6"        &    1.4"     &  19\%       &  9\%           \\ \hline
12.05.99 & $6\times30\rightarrow180$   & $7\times10\rightarrow70$  &  1.3"        &    1.2"     &  52\%       & 27\%           \\ \hline
12.07.99 & $6\times30\rightarrow180$   & $6\times10\rightarrow60$  &  1.3"        &    1.0"     &  68\%       & 37\%           \\ \hline
12.09.99 & $7\times30\rightarrow210$   & $8\times10\rightarrow80$  &  1.5"        &    1.2"     &  14\%       & 38\%           \\ \hline
12.11.99 & $2\times30\rightarrow 60$   & $2\times10\rightarrow20$  &  1.8"        &    1.7"     &   2\%       &  3\%           \\ \hline
12.13.99 & $5\times30\rightarrow150$   & $3\times10\rightarrow30$  &  1.3"        &    1.1"     &  54\%       & 26\%           \\ \hline
\enddata
\end{deluxetable}

\begin{deluxetable}{lll}
\tabletypesize{\scriptsize}
\tablecaption{LMC Dwarf Nova candidate coordinates}
\tablewidth{0pt}
\tablehead{\colhead{Candidate} & \colhead{RA(2000)} & \colhead{DEC(2000)} }
\startdata
      SHZ1     &  5:30:42.60    &  -70:47:32.5          \\ \hline
      SHZ2     &  5:31:38.25    &  -70:38:33.7          \\ \hline      
      SHZ3     &  5:34:42.07    &  -70:49:23.6          \\ \hline      
      SHZ4     &  5:35:10.66    &  -70:28:02.0          \\ \hline      
      SHZ5     &  5:35:47.75    &  -70:24:10.2          \\ \hline      
      SHZ6     &  5:35:24.47    &  -70:23:55.4          \\ \hline      
\enddata
\end{deluxetable}

\begin{deluxetable}{lll}
\tabletypesize{\scriptsize}
\tablecaption{Variable coordinates}
\tablewidth{0pt}
\tablehead{\colhead{Candidate} & \colhead{RA(2000)} & \colhead{DEC(2000)} }
\startdata
      V1     & 5:33:25.77    & -70:52:05.7          \\ \hline
      V2     & 5:30:45.36    & -70:50:42.9          \\ \hline
      V3     & 5:33:28.53    & -70:46:08.2          \\ \hline
      V4     & 5:33:02.97    & -70:35:57.6          \\ \hline
      V5     & 5:32:33.25    & -70:28:48.5          \\ \hline
      V6     & 5:31:01.26    & -70:18:11.6          \\ \hline
      V7     & 5:36:39.89    & -70:37:53.5          \\ \hline
\enddata
\end{deluxetable}

\end{document}